\documentclass[]{mem}
\usepackage{natbib}
\usepackage{txfonts}
\usepackage{balance}
\usepackage{graphicx}
\usepackage[a4paper]{hyperref}

\begin{document}

\title{Evolution of Filament Barbs}

\subtitle{}

\author{Rui Liu, Yan Xu, \& Haimin Wang}

\offprints{R. Liu}

\institute{Space Weather Research Laboratory, Center for Solar-Terrestrial Research, NJIT, Newark,
NJ 07102, USA \email{rui.liu@njit.edu}}

\authorrunning{}

\titlerunning{}

\abstract{We present a selected few cases in which the sense of chirality of filament barbs changed
within as short as hours. We investigate in detail a quiescent filament on 2003 September 10 and
11. Of its four barbs displaying such changes only one overlay a small polarity inversion line
inside the EUV filament channel (EFC). No magnetic elements with magnitude above the noise level
were detected at the endpoints of all barbs. In particular, a pair of barbs first approached toward
and then departed from each other in H$\alpha$, with the barb endpoints migrating as far as
$\sim$10$''$. We conclude that the evolution of the barbs was driven by flux emergence and
cancellation of small bipolar units at the EFC border.

}

\maketitle{}

\section{Introduction}

\begin{figure}[]
\centering \resizebox{0.75\hsize}{!}{\includegraphics[clip=true]{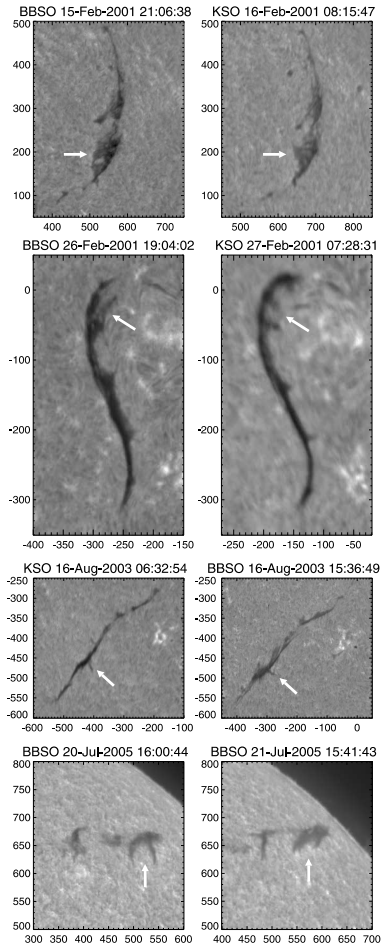}} \caption{
\footnotesize Four events showing the orientation change of filament barbs, as marked by arrows.
From top to bottom, the time difference between the left- and the right-column image are about 11,
12, 9, and 24 hours, respectively.} \label{liu-fig:barbs}
\end{figure}

Barbs are lateral extensions from the filament spine and apparently terminate in the chromosphere.
Similar to the spine, they are composed of parallel, thin threads \citep[e.g.,][]{liu-lin05a}. The
direction in which they extend from the spine reveals the \emph{chirality} of the filament, which
follows a similar hemispheric rule as other solar magnetic structures \citep{liu-martin98}. Martin
and colleagues \citep{liu-me94, liu-mm97} have established empirically that a \emph{dextral}
(\emph{sinistral}) filament has right-bearing (left-bearing) barbs in the same sense as right-hand
(left-hand) ramps off an elevated highway, as which the filament spine is regarded by an analogy.

\citet{liu-me94} suggested that the ends of barbs are fixed at patches of parasitic polarities
(also termed minority polarities), which are opposite in polarity to the surrounding network
elements. Contradictory observations, however, have since been reported on this particular issue.
For example, \citet{liu-vb04} studied a barb terminated between the network elements, where the
field is very weak, with no evidence for flux cancellation. \citet{liu-cmp05} reported that two
barbs of an H$\alpha$ filament terminated over minority-polarity inversion lines.
\citet{liu-lin05a, liu-lin05b} found that the majority of barbs of quiescent filaments end within
the network boundaries as represented by flow converging regions, where the fields are often too
weak to be detected.

Two different 3D models have been proposed on how barbs are magnetically structured. Via
morphological deductions, Martin and colleagues construct an \emph{arcade model} \citep{liu-me94,
liu-mm97}, in which both the spine and the barbs are composed of magnetic arcades aligned with thin
threads, implying that barbs terminate in the parasitic polarity.

Alternatively, Aulanier and colleagues \citep{liu-ad98, liu-aulanier98, liu-aulanier99, liu-asm00}
model the H$\alpha$ absorbing features in the filament channel, i.e., the spine, the barbs, and the
surrounding fibrils, as the dipped portions of field lines (hence named \emph{dip model}). The dips
in the barbs are formed by the perturbations of the background field due to the existence of
parasitic polarities. With projection effects, a continuous pattern of dipped field lines could
give the illusion that barbs are made of vertical fields joining the spine to the photosphere. The
barb orientation (i.e., right- or left-bearing), is not only constrained by the sign of magnetic
helicity, but also by the distribution of parasitic polarities \citep{liu-aulanier99, liu-asm00},
whose evolution drives the evolution of barbs \citep{liu-aulanier99}. \citet{liu-aulanier98} showed
that current dissipation at quasi-separatrix layers are associated with these lateral dips, where
Ca II (8542 {\AA}) brightenings were observed. Using magnetograms as boundary conditions, the
filament structures were successfully reconstructed through a constant-$\alpha$ magnetohydrostatic
extrapolation \citep{liu-aulanier99, liu-asm00}. The dip model has been further developed using
non-linear force-free solutions \citep[e.g.,][]{liu-vb04}.

Observation and modeling have been advanced to EUV, in which a filament is observed as a much wider
intensity-depletion region than its H$\alpha$ counterpart \citep{liu-hst01}. \citet{liu-wang01}
reported that He II (304 {\AA}) filaments are rooted in areas of on-going flux cancellation, and
that those legs and barbs decay away after the cancellation. \citet{liu-as02} found that the EUV
filament channel (EFC) is dominated by extensions with similar morphology to H$\alpha$ barbs. In
their data-driven modeling, H$\alpha$ structures are matched by high-altitude dips while the EFC by
low-altitude ones. They identified the endpoints of H$\alpha$ barbs with weak parasitic polarities
(3--10 G), while magnetic concentrations, shown as isolated bright patches in EUV, are at the outer
parts of the EFC.

In this paper, we present observations of the dynamic evolution of filament barbs (Section 2), in
which the sense of chirality, as defined in \citet{liu-martin98}, can be changed within 24 hours.
Concluding remarks are made in Section 3.

\section{Observation and Analysis}

\begin{figure}[]
\centering \resizebox{0.9\hsize}{!}{\includegraphics[clip=true]{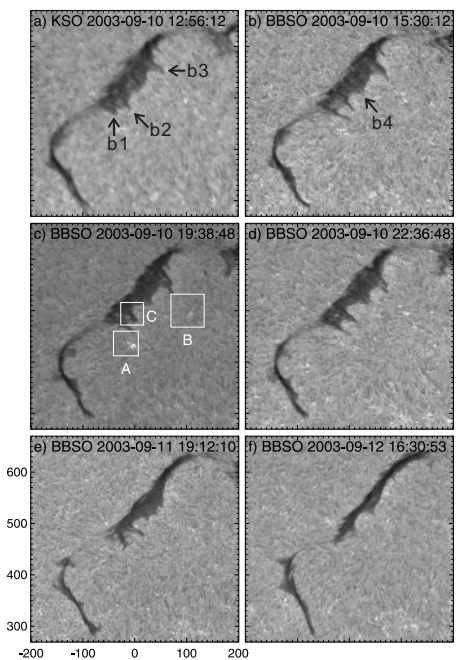}}
\caption{\footnotesize Snapshots of KSO and BBSO H$\alpha$ images on 2003 September 10 and 11. }
\label{liu-fig:barb}
\end{figure}

From daily H$\alpha$ images obtained by Kanzelh\"{o}he Solar Observatory (KSO) and Big Bear Solar
Observatory (BBSO), a selected few examples are shown in Fig.~\ref{liu-fig:barbs}, in which
remarkable changes of the barb orientation occurred within 9--24 hours, which cannot be explained
by the the change of perspective only. The first three cases show that the barbs marked by arrows
changed from left- to right-bearing, or vice versa. The last one shows that two barbs changed from
apparently perpendicular to the spine to left-bearing.

Below we investigated in detail a quiescent filament in the northern hemisphere during 2003
September 10 and 11. In Fig.~\ref{liu-fig:barb}, snapshots of H$\alpha$ images show the highly
dynamic and complicated evolution of the filament barbs. All images are registered to the one in
Panel (e), in which the barbs were close to the disk center. In Fig.~\ref{liu-fig:barb}(a), marked
by arrows are three barbs apparently perpendicular to the spine. As time progressed, $b1$ and $b2$
extended their lengths, and developed into a right-bearing and a left-bearing barb, respectively
(Fig.~\ref{liu-fig:barb}(b)). Meanwhile, a new barb, $b4$, was forming to the southeast of $b3$
(Fig.~\ref{liu-fig:barb}(b)), while $b1$ began to ``rotate'' clockwise and $b2$ counterclockwise.
As of 2003 September 11 19:12 UT (Fig.~\ref{liu-fig:barb}(e)), $b1$ has changed from right-bearing
to left-bearing, opposite to the evolution of $b2$, while both $b3$ and $b4$ have developed into
left-bearing barbs. On the next day (2003 September 12), all four barbs almost disappeared
(Fig.~\ref{liu-fig:barb}(i)).

\begin{figure}[]
\centering \resizebox{0.85\hsize}{!}{\includegraphics[clip=true]{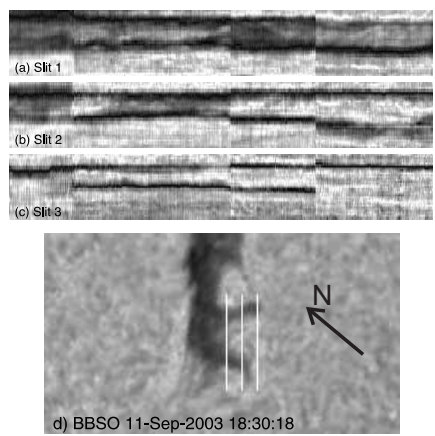}} \caption{
\footnotesize H$\alpha$ slit images. Slits 1, 2, and 3 are positioned from left to right,
respectively, as shown in the bottom panel. The arrow points to the north.} \label{liu-fig:slits}
\end{figure}

Particularly interesting is the evolution of $b1$ and $b2$. In Fig.~\ref{liu-fig:slits}(d), they
are rotated to a ``horizontal'' position. Before that, we have aligned and then de-stretched
H$\alpha$ images to correct image jitter and warping. We then put three slits across $b1$ and $b2$
(Fig.~\ref{liu-fig:slits}(d)), referred to as Slits 1, 2, and 3, from left to right, respectively.
Slices of H$\alpha$ images cut by the slits are shown in Fig.~\ref{liu-fig:slits}(a)--(c), in which
$b1$ (lower streak) and $b2$ (upper streak) first approached toward, and then departed from, each
other, during the two-day interval. Note that the discontinuities are due to data gaps (see also
Fig.~\ref{liu-fig:flux}).

\begin{figure}[]
\centering \resizebox{0.85\hsize}{!}{\includegraphics[clip=true]{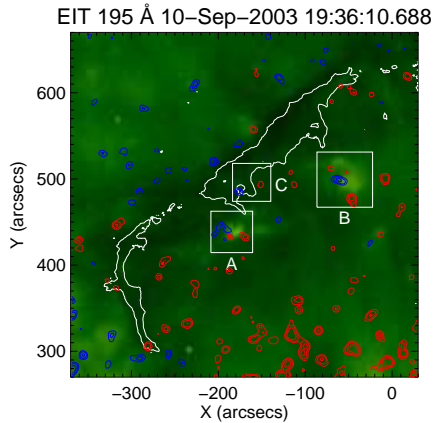}} \caption{
\footnotesize EIT 195 {\AA} image overlaid by H$\alpha$ filament (white contours) and MDI
magnetogram. Contour levels of fields are 30, 60, and 120 G for positive polarities (red), and
-120, -60, and -30 G for negative polarities (blue). } \label{liu-fig:over}
\end{figure}

Two H$\alpha$ brightening regions nearby the barbs of interest are identified in
Fig.~\ref{liu-fig:barb}(d) (labelled A and B), with corresponding brightening in SOHO/EIT 195 {\AA}
(see Fig.~\ref{liu-fig:over}). When overlaying the H$\alpha$ filament and SOHO/MDI magnetograms on
corresponding EIT 195 {\AA} images (Fig.~\ref{liu-fig:over}), one can see that both brightening
regions are co-spatial with dipolar units composed of a magnetic element of parasitic (negative)
polarity in close contact with magnetic element(s) of major (positive) polarity. A third dipolar
unit which is spatially associated with $b2$ is denoted as C (Fig.~\ref{liu-fig:barb}(d) and
Fig.~\ref{liu-fig:over}).

Integrating MDI flux over the three rectangular regions, inside which the dipolar units are
localized over the two-day interval, one can see that significant flux emergence and cancellation
of minor polarity occur in both Regions A and B, while there is minimal activity in Region C
(Fig.~\ref{liu-fig:flux}(b) and (c)). Interestingly, the profile of MDI fluxes bears a remarkable
similarity to that of EIT data counts in the same region (Fig.~\ref{liu-fig:flux}), suggestive of
the connection of the two phenomena.

\begin{figure}[]
\resizebox{\hsize}{!}{\includegraphics[clip=true]{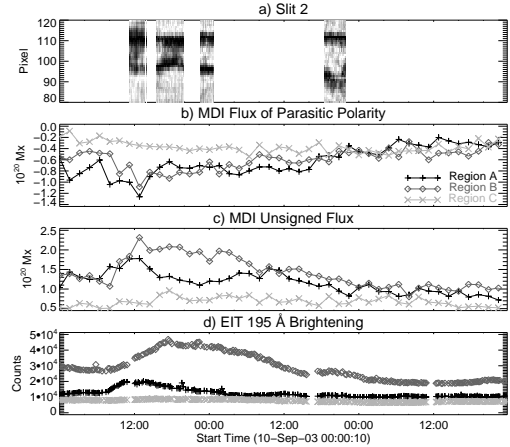}} \caption{ \footnotesize Profiles of
MDI fluxes and EIT 195 {\AA} data counts integrating over the regions specified in
Fig.~\ref{liu-fig:over}, along with slices of H$\alpha$ images cut by Slit 2. }
\label{liu-fig:flux}
\end{figure}

\section{Conclusion}
Of the four barbs displaying obvious changes of orientation on 2003 September 10 and 11, $b2$ is
the only one overlying a small polarity inversion line inside the EFC (Fig.~\ref{liu-fig:over}).
The endpoints of the rest three barbs are not fixed at any magnetic element with magnitude above
the noise level. It is unlikely that changes of such weak fields would result in the ``migration''
of the barb ends as far as $\sim$10$''$ (see Fig.~\ref{liu-fig:flux}(a)). It is more intuitive,
however, to conclude that the barbs are not rooted in the photosphere, but their evolution was
associated with flux emergence and cancellation of the dipole units at the EFC border. This casts
doubt on the reliability of barbs as the indicator of the filament chirality.

\begin{acknowledgements}
H$\alpha$ data are provided through the Global High Resolution H-alpha Network. This work was
supported by NASA grant NNX08-AJ23G and NNX08-AQ90G, and by NSF grant ATM-0849453 and ATM-0839216.
\end{acknowledgements}

\bibliographystyle{aa}

\end{document}